\documentclass{ws-procs9x6}

\begin{document}

\title{The Distribution of Quasars and Galaxies in
            Radio Color-Color and Morphology Diagrams   }

\author{\v{Z}. IVEZI\'{C}, R.J. SIVERD, W. STEINHARDT, A.S. JAGODA,
G.R. KNAPP, R.H. LUPTON, D. SCHLEGEL, P.B. HALL, G.T. RICHARDS,
J.E. GUNN, M.A. STRAUSS, M. JURI\'{C}, AND P. WIITA}

\address{Department of Astrophysical Sciences \\
Princeton University, Princeton, NJ 08544, USA}

\author{M. GA\'{C}E\v{S}A AND V. SMOL\v{C}I\'{C}}

\address{Department of Physics \\
University of Zagreb, Bijeni\v{c}ka Cesta 4, Zagreb, Croatia}

\maketitle

\abstracts{
We positionally match the 6 cm GB6, 20 cm FIRST and
NVSS, and 92 cm WENSS radio catalogs and find 16,500 matches in
$\sim$3,000 deg$^2$ of sky. Using this unified radio database,
we construct radio ``color-magnitude-morphology" diagrams and find
that they display a clear structure, rather than a random scatter.
We propose a simple, yet powerful, method for morphological classification
of radio sources based on FIRST and NVSS measurements. For a subset of
matched sources, we find optical identifications using the SDSS Data
Release 1 catalogs, and separate them into quasars and galaxies.
Compact radio sources with
flat radio spectra are dominated by quasars, while compact sources
with steep spectra, and resolved radio sources, contain substantial
numbers of both quasars and galaxies.}

\section{           The Era of Modern Radio Surveys }

Statistical studies of the radio emission from extragalactic sources
are entering a new era due to the availability of large sky area
high-resolution radio surveys that are sensitive to mJy levels
(e.g. FIRST, Becker, White \& Helfand 1995; GB6, Gregory et al. 1996;
WENSS, Rengelink et al. 1997; NVSS, Condon et al. 1998). The catalogs based
on these surveys contain millions of sources, have high completeness and
low contamination, and are available in 
digital form. The wide wavelength region spanned by these surveys,
from 6 cm for GB6 to 92 cm for WENSS, and the detailed morphological information
at 20 cm provided by FIRST and NVSS, allow significant quantitative and
qualitative advances in studies of radio sources.

Here we present a preliminary analysis of sources detected by the GB6, NVSS,
FIRST and WENSS radio surveys, and of the subset also
detected by the optical Sloan Digital Sky Survey (SDSS; York et al. 2000;
Stoughton et al. 2002, and references therein). Matching FIRST and NVSS
allows a robust estimate of the source morphology at 20 cm,
and the addition of GB6 and WENSS data allows the determination of
the radio spectral slope and curvature.  SDSS identifications enable
the separation of sources into quasars and galaxies, and, for some objects,
also provides the redshifts.

\begin{figure}[t]
\vskip -0.2in
{\hskip 1.0in \epsfxsize=4.5in\epsfbox{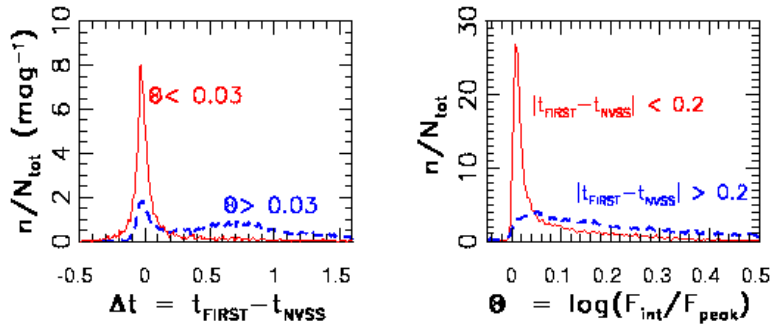}}
\caption{The bimodal distribution of the $\Delta t = t_{FIRST}-t_{NVSS}$ 
magnitude differences (left). This bimodality essentially reflects the 
separation between isolated core sources ($|\Delta t| < 0.2$) and lobe 
and complex sources ($|\Delta t|> 0.2$). We confirmed this conclusion by 
visually inspecting about 1000 $2'\times 2'$ FIRST images (Ivezi\'{c} 
et al. 2004, in prep). The majority of lobe and complex sources, and some
core sources, are resolved by FIRST on 5 arcsec scale ($\Theta > 0.03$, 
right panel).}
\end{figure}

\section{ The Cross-identification of Radio Sources}

The cross-identification of radio surveys at different wavelengths, and with
different resolutions, is not straightforward (e.g. Becker et al. 1995;
Ivezi\'{c} et al. 2002). 
However, the accurate positions provided by FIRST allow simple
positional matching with high completeness and a low random contamination rate.
Based on an analysis of positional differences, we adopted maximum positional
discrepancies between FIRST and the other three surveys of 15$''$,
30$''$ and 60$''$, for NVSS, WENSS and GB6, respectively. This choice
yields completeness of over 80\% and a false matching rate below 1\%.

The current positional overlap of the GB6, FIRST, NVSS, and WENSS catalogs
includes 16,500 sources in about 3,000 deg$^2$. The matched sample is flux
limited by GB6 for sources with spectral slope $\alpha < 0$, and by WENSS for
sources with $\alpha > 0$, where $F_\nu \propto \nu^\alpha$. This is the
largest database of the radio spectral and morphological measurements assembled
to date.

For convenience, we express all fluxes on the AB$_\nu$ magnitude system of
Oke \& Gunn (1983), where $m = -2.5 \log ({F_\nu / 3631 \, {\rm Jy}})$.
The survey limits expressed in this system are $s < 13.3$, $t_{FIRST} < 16.4$,
$t_{NVSS} < 15.4$, and $n < 13.3$, for GB6, FIRST, NVSS, and WENSS, respectively
(magnitudes are named by the first letter of the corresponding wavelength
in cm).

\begin{figure}[t]
\vskip -0.2in
\rightline{\hskip 0.5in \epsfxsize=2.2in\epsfbox{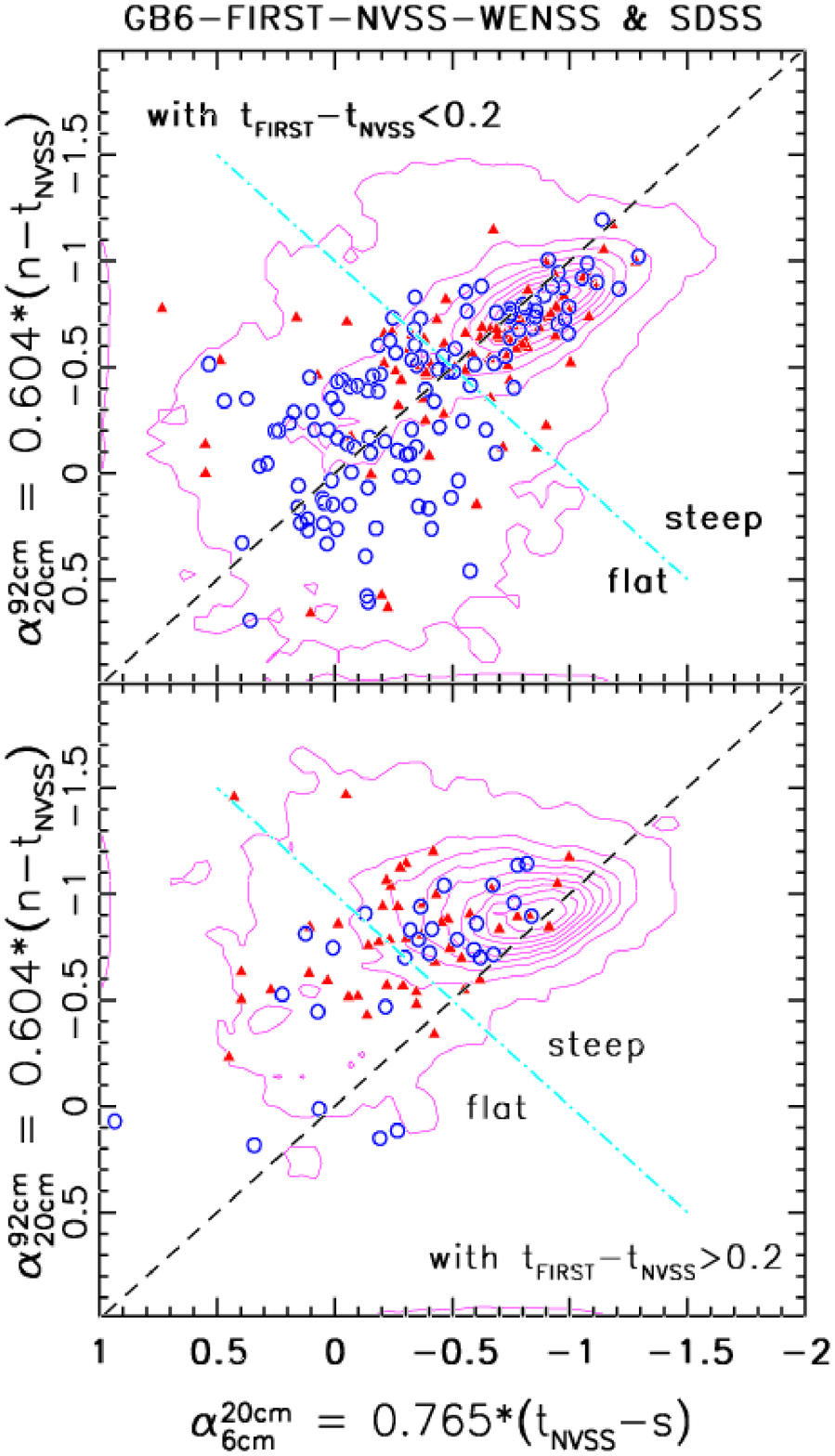}}
\vskip -3.8in
\leftline{\hskip 0.0in \epsfxsize=2.2in\epsfbox{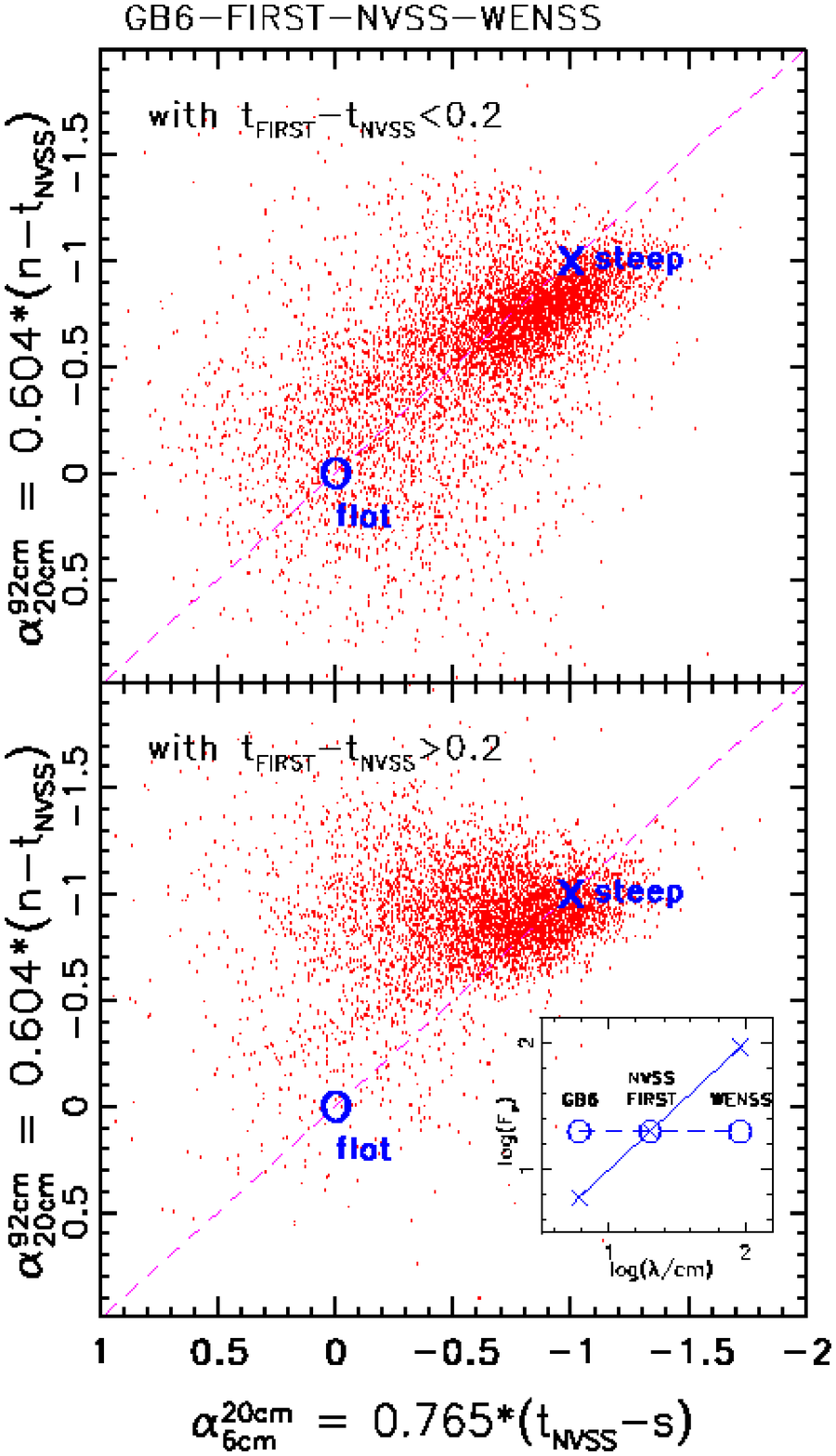}}
\caption{The distribution of radio sources detected by GB6, FIRST,
NVSS, and WENSS in the 6-20-92 cm radio spectral slope
diagram is shown in the two left panels for ``core'' (top) and
``lobe'' (bottom) sources, separated using the flux difference between
FIRST and NVSS measurements. The diagonal dashed lines show the $y=x$
locus, with the positions for $\alpha = 0$ and $\alpha = -1$ marked
as ``flat'' and ``steep'', respectively. The SED difference between
these two cases is illustrated in the insert in the bottom left panel.
The two right panels compare the distributions of all matched radio
sources (contours), to those for sources detected by the SDSS
(triangles for galaxies and circles for quasars).}
\end{figure}

\vskip -0.1in
\section{ Radio Morphology from FIRST and NVSS Data}

FIRST is a high resolution (5$''$) interferometric survey and thus
over-resolves large sources relative to the lower resolution
(45$''$) NVSS survey, leading to underestimated fluxes.  Comparing the
two surveys efficiently separates point and extended sources.
The $\Delta t = t_{FIRST}-t_{NVSS}$ magnitude differences have a
bimodal distribution (Figure 1) which reflects the separation 
between isolated core sources ($|\Delta t| < 0.2$) 
and lobe and complex sources ($|\Delta t|> 0.2$). 
In addition to $\Delta t$, another useful morphological parameter
is the ratio of integrated and peak FIRST fluxes --
sources with $\theta=\log(F_{int}/F_{peak}) < 0.03$ appear unresolved by FIRST
($\Delta t$ and $\theta$ are not equivalent because they measure size
on different scales -- 5$''$ and 45$''$).

\section{             Radio Color-Color Diagrams and Optical Identifications }

The left panels in Figure 2 show the distributions of core (top)
and lobe (bottom) sources (separated using $t_{FIRST}-t_{NVSS}$) in
the 6-20-92 cm radio spectral slope (i.e. ``color''-``color'') diagram.
Note that there are more flat-spectrum sources in the top panel. According
to size measurements from the FIRST survey and Figure 1, the overwhelming
majority of flat-spectrum sources are also unresolved on 5$''$ scales
($\theta < 0.03$) and on 45$''$ scales.
%
%
The right panels in Figure 2 compare the distributions of
all matched radio sources (contours) to those for sources optically
identified by SDSS. The compact flat-spectrum sources are dominated by
quasars, while compact sources with steep spectra, and resolved radio
sources, include substantial numbers of both quasars and galaxies.
It is also noteworthy that the optically identified sources are {\em not}
representative of the whole radio population.

\section{                  Conclusions                  }

The distribution of sources in radio
``color-magnitude-morphology" space displays clear structure which
encodes detailed information about the astrophysical processes that are
responsible for the observed radio emission, and thus provides strong
constraints for the models of these processes.

\vskip -0.1in
\phantom{x}
\section*{Acknowledgments}
\footnotesize{Funding for the creation \& distribution of the SDSS Archive
(http://www.sdss.org/)~has been provided by
the Alfred P. Sloan Foundation, the Participating Institutions, the National Aeronautics
and Space Administration, the National Science Foundation, the U.S. Department of Energy,
the Japanese Monbukagakusho, and the Max Planck Society.
}
\vskip -0.3in
\phantom{x}

\end{document}